\documentstyle[12pt,epsf]{article}

\advance\hoffset by -1.1truecm
\advance\voffset by -1.0truecm

\begin{document}

\title{Hilbert Space Representation of the Minimal Length 
Uncertainty Relation}
\author{Achim Kempf\thanks{supported by Studienstiftung des Deutschen 
Volkes, BASF-fellow, 
a.kempf@amtp.cam.ac.uk} , Gianpiero Mangano\thanks{
supported by University of Naples, {\it Federico II}, \rm Italy, 
g.mangano@amtp.cam.ac.uk}, Robert B. Mann\thanks{
supported by NSERC of Canada, rbm20@amtp.cam.ac.uk}\\ \\
Department of Applied Mathematics \& Theoretical Physics\\
University of Cambridge\\ 
Cambridge CB3 9EW, U.K.}

\date{}
\maketitle

\begin{abstract}
The existence of a minimal observable length has long been suggested  
in quantum gravity  as well as in string theory. In this context
a generalized uncertainty relation has been derived which quantum 
theoretically describes the minimal length as a minimal uncertainty 
in position measurements. Here we study in full detail
the quantum mechanical structure which underlies this uncertainty relation.
\end{abstract}

\vskip-13truecm
\hskip1.5truecm
{\tt DAMTP/94-105, hep-th/9412167, and 
 Phys.Rev.{\bf D52}:1108 (1995)\rm  
\vskip14truecm

\newpage

\section{Introduction}
One of the major problems in quantum gravity 
is that the introduction of gravity into quantum field theories appears 
to spoil their renormalizability. On the other hand it has long been
suggested that gravity itself should lead to an effective cutoff in the
ultraviolet, {\it i.e.} to a minimal observable length. 
The argument is based on the
expectation that the high energies used in trying to resolve small distances 
will eventually significantly disturb the spacetime structure by their 
gravitational effects. 
While conventional spacetime locality seems to be probed down to scales
of about $1$ TeV \cite{khuri} it is quite clear that this type of
effect ought to occur at least at energy scales as large as
the Planck scale. 
If indeed gravity induces a lower bound to the 
possible resolution of distances, gravity could in fact be expected to
regularize quantum field theories rather than rendering them 
nonrenormalizable. 
It is a natural, though non-trivial assumption
that a minimal length should quantum theoretically
be described as a nonzero minimal uncertainty $\Delta x_0$ 
in position measurements. String-theoretic arguments also 
lead to a minimal length effectively of the form of a minimal 
position uncertainty. See {\it e.g.}
 \cite{townsend}-\cite{maggiore} and
for a recent survey \cite{garay}. 

The purpose of this paper is to
develop a generalized quantum theoretical framework which implements
the appearance of a nonzero minimal uncertainty in positions. We will
here confine ourselves to exploring the implications of such a minimal
uncertainty in the context of non-relativistic
quantum mechanics. A further paper
extending the framework to field theory is in preparation.

Our analysis is motivated by the results of 
\cite{ak-lmp-bf}-\cite{ak-np2} 
where the more general case, which includes nonzero minimal uncertainties in
momenta as well as position, was considered. This general case is far more 
difficult to handle, since (as we will 
explain) neither a position nor a momentum space representation is viable. 
Instead one  has to resort to a
generalized Bargmann-Fock space representation. The construction of 
generalized Bargmann-Fock Hilbert spaces was realized by making 
use of algebraic techniques developed in the field
 of quantum groups; for a comparison
 see {\it e.g.} \cite{FRT}-\cite{sm-thesis}. 
However, using the discrete Bargmann Fock basis, actual calculations
typically involve finite difference equations and infinite sums
rather than differential equations and integrals. For these technical
reasons it has so far only in examples been possible to prove {\it e.g.} 
that  ultraviolet divergences may be  regulated using this approach. 
In the present paper we consider only the 
case of minimal uncertainty in position, taking the
minimal uncertainty in momentum to vanish. This case is of interest
because there still exists a continuous 
momentum space representation. Hence it allows us to explore the physical
implications of a minimal length in a manner that is technically 
much easier to handle than the general case.

\section{Minimal length uncertainty relations}

In one dimension the simplest generalized uncertainty relation 
which implies the appearance of a nonzero minimal uncertainty 
$\Delta x_0$ in position has the form:
\begin{equation}
\Delta x \Delta p \ge \frac{\hbar}{2} (1 + \beta (\Delta p)^2 + \gamma)
\label{ucra}
\end{equation}
where $\beta$ and $\gamma$ are positive and independent 
of $\Delta x$ and $\Delta p$ (but may in general
depend on the expectation values of ${\bf{x}}$ and ${\bf{p}}$). The curve of
minimal uncertainty is illustrated in Fig. 1. 
\vskip0.5truecm
\epsfysize=2.9in \centerline{\epsfbox{pic1.ps}}
%
%
%
%
\centerline{\small \it Fig1: Modified uncertainty relation, 
implying a `minimal length' $\Delta x_0>0$ }
\vskip0.4truecm
While in ordinary quantum mechanics $\Delta x$
can be made arbitrarily small by letting $\Delta p$ grow correspondingly,
this is no longer the case if (\ref{ucra}) holds. If for 
decreasing $\Delta x$ $\Delta p$ 
increases, the new term $\beta (\Delta p)^2$ on the rhs of
(\ref{ucra}) 
will eventually grow faster than the lhs. 
Hence $\Delta x$ can no longer be made arbitrarily small.

This type of generalized uncertainty relation has appeared in the context 
of quantum gravity and string theory, see {\it e.g.}  \cite{maggiore}
and independently from formal considerations in ref. \cite{ak-jmp-ucr}.
It allows one to
express the (non-trivial) idea that
a minimal length should quantum theoretically be described as a minimal 
uncertainty in position measurements. 
\smallskip\newline
More generally, the relation
\begin{equation}
\Delta x \Delta p \ge \frac{\hbar}{2} (1 + \alpha (\Delta x)^2 + 
\beta (\Delta p)^2 + \gamma)
\label{ucrab}
\end{equation}
leads to a nonzero minimal uncertainty in both position $\Delta x_0$
and momentum  $\Delta p_0$ for $\alpha > 0$. 
\smallskip\newline
Now in general it is known that for any pair of 
observables $A,B$ which are represented as 
symmetric operators on a domain of $A^2$ 
and $B^2$ the uncertainty relation
\begin{equation}
\Delta A \Delta B \ge \frac{\hbar}{2} \vert \langle [A,B] \rangle \vert
\end{equation}
will hold.  In particular, we see that  
commutation relations of the form
\begin{equation}
[{\bf{x}},{\bf{p}}] = i\hbar (1 + \alpha {\bf{x}}^2 + \beta {\bf{p}}^2)
\label{1dcom}
\end{equation}
underly the uncertainty relation (\ref{ucrab}) with $\gamma =
\alpha \langle {\bf{x}}\rangle^2 + \beta \langle {\bf{p}}\rangle^2$.

\section{Hilbert space representation}

We will now construct a Hilbert space representation of such a
commutation relation. Let us clarify that we generally 
require physical states
not only to be normalizable, but to also have
well defined expectation values of position and momentum and also 
well defined uncertainties in these quantities.
It is important to note that this implies that physical states
always lie in the common domain $D_{{\bf{x}},{\bf{x}}^2,{\bf{p}},{\bf{p}}^2}$ 
of the symmetric operators ${\bf{x}},{\bf{p}},{\bf{x}}^2,{\bf{p}}^2$. 
On $D_{{\bf{x}},{\bf{x}}^2,{\bf{p}},{\bf{p}}^2}$ the uncertainty relation holds
from which we can already conclude that physical states are constrained
to the `allowed' region of Fig. 1. 
Actually we can further derive from the uncertainty relation a
severe contraint on the possible ansatzes for
 Hilbert space representations. 

\subsection{Representation theoretic 
consequences of the uncertainty relations}

In ordinary quantum mechanics
${\bf{x}}$ and ${\bf{p}}$ could {\it e.g.} be represented as multiplication 
or differentiation
operators acting on square integrable position- or momentum- space wave 
functions $\psi(x):= \langle x \vert \psi \rangle $ 
or $\psi(p):= \langle p \vert \psi \rangle $, where
the $\vert x \rangle$ and $\vert p \rangle$ are position and momentum `eigenstates'.
Strictly speaking the $\vert x \rangle$ and $\vert p \rangle$ are not physical states since
they are  not normalizable and thus not in the Hilbert space.
However, the operators ${\bf{x}}$ and ${\bf{p}}$ are essentially self-adjoint and the
`eigenstates' can be approximated to arbitrary precision 
by sequences $\vert \psi_n \rangle $ of physical states of increasing
localization in position or momentum space: 
$$\lim_{n\rightarrow\infty}\Delta x_{\vert \psi_n \rangle} = 0$$ or
$$\lim_{n\rightarrow\infty}\Delta p_{\vert \psi_n \rangle} = 0$$
As has been pointed out in \cite{ak-jmp-ucr,ak-np2}
this situation changes drastically with the introduction of 
minimal uncertainties $\Delta x_0 \ge 0$ and/or $\Delta p_0 \ge 0$.
For example a nonzero minimal uncertainty in position 
\begin{equation}
(\Delta x)^2_{\vert \psi \rangle} = \langle \psi \vert ({\bf{x}} - \langle \psi\vert 
{\bf{x}} \vert \psi \rangle )^2 \vert \psi \rangle
\ge \Delta x_0 \qquad \forall \quad\vert \psi \rangle 
\end{equation}
implies that there cannot be any physical state which is a 
position eigenstate 
since an eigenstate would of course have zero
uncertainty in position. 
\smallskip\newline
Of course this does not exclude the 
existence of unphysical, `formal
position eigenvectors' which lie in the domain of ${\bf{x}}$ alone but
not in $D_{{\bf{x}},{\bf{x}}^2,{\bf{p}},{\bf{p}}^2}$. As we will see, such formal ${\bf{x}}$-eigenvectors
do exist and are of infinite energy.
Most importantly however, unlike in ordinary quantum mechanics,
it is no longer possible to approximate
formal eigenvectors through a sequence of physical states of
uncertainty in positions decreasing to zero. 
This is because now all physical states have at least 
a finite minimal uncertainty in position. 
\smallskip\newline
Technically, as we will see, a minimal uncertainty 
in position will mean that the
position operator is no longer essentially self-adjoint but
only symmetric. While the preservation of the 
symmetry insures that all expectation values
are real, giving up self-adjointness opens the way for the introduction of
minimal uncertainties. 
\smallskip\newline
However, since there are then no more position eigenstates $\vert x \rangle$ in the
representation of the Heisenberg algebra, the Heisenberg algebra 
will no longer find a Hilbert space representation on position wave 
functions $\langle x \vert \psi \rangle$. 
Similarly a minimal uncertainty in momentum also
abandons momentum space wave functions. In this general case we therefore
resorted to a generalized Bargmann Fock 
representation \cite{ak-jmp-ucr,ak-np2}.
\smallskip\newline
Here we will restrict ourselves to the case $\alpha = 0$ where
there is no minimal uncertainty in momentum. This will allows us to work with
the convenient representation of the commutation relations on momentum
space wave functions.
\smallskip\newline
As we will see, the states of maximal 
localization will be proper physical states. We can use 
them to define a `quasi-position' representation. This representation has a 
direct interpretation in terms of position measurements, although it
does of course not diagonalise ${\bf{x}}$. 

\subsection{Representation on momentum space}
We consider the associative Heisenberg algebra generated 
by ${\bf{x}}$ and ${\bf{p}}$ obeying the commutation relation ($\beta >0$)
\begin{equation}
[{\bf{x}},{\bf{p}}] = i\hbar (1 + \beta {\bf{p}}^2)
\label{cr1dim}
\end{equation}
The corresponding uncertainty relation is
\begin{equation}
\Delta x \Delta p \ge \frac{\hbar}{2} (1+\beta (\Delta p)^2 
+ \beta \langle {\bf{p}} \rangle^2)
\end{equation}
with the curve on the boundary of the allowed 
region being (see Fig.1):
\begin{equation}
\Delta p = \frac{\Delta x}{\hbar\beta}  \pm 
\sqrt{\left(\frac{\Delta x}{\hbar \beta}\right)^2 -\frac{1}{\beta} 
- \langle{\bf{p}}\rangle^2}
\end{equation}
One reads off the minimal position uncertainty
\begin{equation}
\Delta x_{min}(\langle{\bf{p}}\rangle) 
= \hbar \sqrt{\beta} \sqrt{1+\beta \langle{\bf{p}}\rangle^2}
\label{xmin}
\end{equation}
so that the absolutely smallest uncertainty in positions has the value
\begin{equation}
\Delta x_0 = \hbar\sqrt{\beta}
\end{equation}
There is no nonvanishing minimal uncertainty in momentum. In fact
the Heisenberg algebra can be represented on momentum space
wave functions $\psi(p):=\langle p \vert \psi \rangle $. 
\smallskip\newline
We let ${\bf{p}}$ and ${\bf{x}}$ act as operators
\begin{eqnarray}
{\bf{p}}.\psi(p) & = & p \psi(p) \label{one} \\
{\bf{x}}.\psi(p) & = & i \hbar(1+\beta p^2) \partial_p \psi(p) \label{two}
\end{eqnarray}
on the dense domain $S_{\infty}$ of functions decaying faster than any power.
This representation is easily seen to respect 
the commutation relation (\ref{cr1dim}). Note that, although the
generalization for commutation relations of the type $[{\bf{x}},{\bf{p}}] = i\hbar
f({\bf{p}})$ seems obvious, such generalizations are representation-theoretically
nontrivial, in particular 
if $f$ is not strictly positive. This will be studied elsewhere.

Further, ${\bf{x}}$ and ${\bf{p}}$ are symmetric on the domain $S_{\infty}$
\begin{equation}
\left(\langle\psi\vert{\bf{p}}\right)\vert \phi\rangle = 
\langle\psi\vert\left({\bf{p}}\vert\phi\rangle\right)
 \mbox{ \qquad and \qquad}
\left(\langle\psi\vert {\bf{x}}\right)\vert \phi\rangle = 
\langle\psi\vert\left({\bf{x}}\vert\phi\rangle\right)
\end{equation}
but now with respect to the scalar product:
\begin{equation}
\langle \psi \vert \phi \rangle  =  \int_{-\infty}^{+\infty} 
\frac{dp}{1+ \beta p^2} \psi^{*}(p) \phi(p)
\label{sp}
\end{equation}
The symmetry of ${\bf{p}}$ is obvious. The symmetry of
${\bf{x}}$  can be seen by performing a partial integration. 
\begin{equation}
{\int_{-\infty}^{+\infty}\frac{dp}{1+ \beta p^2}}
 \psi^*(p) i\hbar(1+\beta p^2)\partial_p \phi(p)
= 
{\int_{-\infty}^{+\infty}\frac{dp}{1+ \beta p^2}}
 \left(i\hbar(1+\beta p^2)\partial_p\psi(p)\right)^* \phi(p)
\end{equation}
Thereby the $(1+\beta p^2)^{-1}$ -factor of the measure on
momentum space is needed to cancel a corresponding factor of
the operator representation of ${\bf{x}}$.
\smallskip\newline
The identity operator can thus be expanded as
\begin{equation}
1  =  
{\int_{-\infty}^{+\infty}\frac{dp}{1+ \beta p^2}}
 \vert p \rangle\langle p \vert 
\end{equation}
and the scalar product of momentum eigenstates is therefore:
\begin{equation}
\langle p \vert p^{\prime}\rangle
 = (1+\beta p^2) \delta(p-p^{\prime}) \label{sppp}
\end{equation}
While the momentum operator still is essentially self-adjoint,
the functional analysis of the position operator, as expected 
from the appearance of the minimal uncertainty in positions, changes.

\subsection{Functional analysis of the position operator}
The eigenvalue problem for the position operator takes, on momentum space,
the form of the differential equation:
\begin{equation}
i\hbar (1+\beta p^2)\partial_p \psi_{\lambda}(p) = \lambda 
\psi_{\lambda}(p)
\end{equation}
It can be solved to obtain formal position eigenvectors:
\begin{equation}
\psi_{\lambda}(p) = c {\mbox{ }}
e^{-i\frac{\lambda}{\hbar\sqrt{\beta}} \tan^{-1}(\sqrt{\beta} p)}
\end{equation}
They are normalizable:
\begin{equation}
1 = c c^* \int_{-\infty}^{+\infty} \frac{1}{1+\beta p^2} = c c^* 
\pi/\sqrt{\beta}
\end{equation}
Thus
\begin{equation}
\psi_{\lambda}(p) = \sqrt{\frac{\sqrt{\beta}}{ \pi}} {\mbox{ }}
e^{-i\frac{\lambda}{\hbar\sqrt{\beta}} \tan^{-1}(\sqrt{\beta} p)}
\end{equation}
We know however from the uncertainty
relation that these formal eigenvectors are not physical states.
Let us first proceed with a formal analysis.
\smallskip\newline
We note that there is exactly one eigenvector 
to each of the
eigenvalues $ \pm i$ and that they are normalizable.
Technically this means the following:
The operator ${\bf{x}}^{**}$ which is the bi-adjoint of the 
densely defined symmetric operator ${\bf{x}}$ is symmetric
 and closed and has
 non-empty deficency subspaces \cite{ak-jmp-ucr}:
\begin{equation}
L^{ \perp}_{ \pm i,{\bf{x}}^{**}} := ker({\bf{x}}^* \mp i).D_{{\bf{x}}^*}
\end{equation}
Here we used that ${\bf{x}}^{***} = {\bf{x}}^{*}$ which holds since ${\bf{x}}^{*}$
is closed and defined on a dense domain. From the dimensionalities 
of these subspaces, 
{\it i.e.} since the deficency
indices are (1,1), one concludes that the position operator
is no longer essentially self-adjoint but has
a one-parameter family of self-adjoint extensions instead. 
\smallskip\newline
This functional analytic structure had already been found and
studied in detail
in \cite{ak-jmp-ucr} where the more general case including also
a minimal uncertainty in momentum has been covered.
\smallskip\newline
It is however one of the advantages of our specialization to the case
without a minimal uncertainty in momenta, that we can now construct
the one-parameter family of
diagonalisations of ${\bf{x}}$ explicitly.
To this end we calculate the scalar product of the formal position
eigenstates $\vert \psi_{\lambda} \rangle $ (see Fig.2)
\begin{eqnarray}
\langle \psi_{\lambda^\prime} \vert \psi_{\lambda} \rangle & = &
\frac{\sqrt{\beta}}{\pi} \int_{-\infty}^{+\infty} \frac{dp}{1+\beta p^2}
{\mbox{ }} 
e^{-i\frac{(\lambda-\lambda^\prime)}{\hbar\sqrt{\beta}} 
\tan^{-1}(\sqrt{\beta} p)} \nonumber \\
 \nonumber \\ 
  & = & \frac{2 \hbar\sqrt{\beta}}{
\pi (\lambda - \lambda^{\prime})} {\mbox{ }} \sin\left(\frac{\lambda 
-\lambda^{\prime}}{2\hbar \sqrt{\beta}} 
\pi\right)
\label{spfp}
\end{eqnarray}
\vskip-4.0truecm
\epsfxsize=3.5in \centerline{\epsfbox{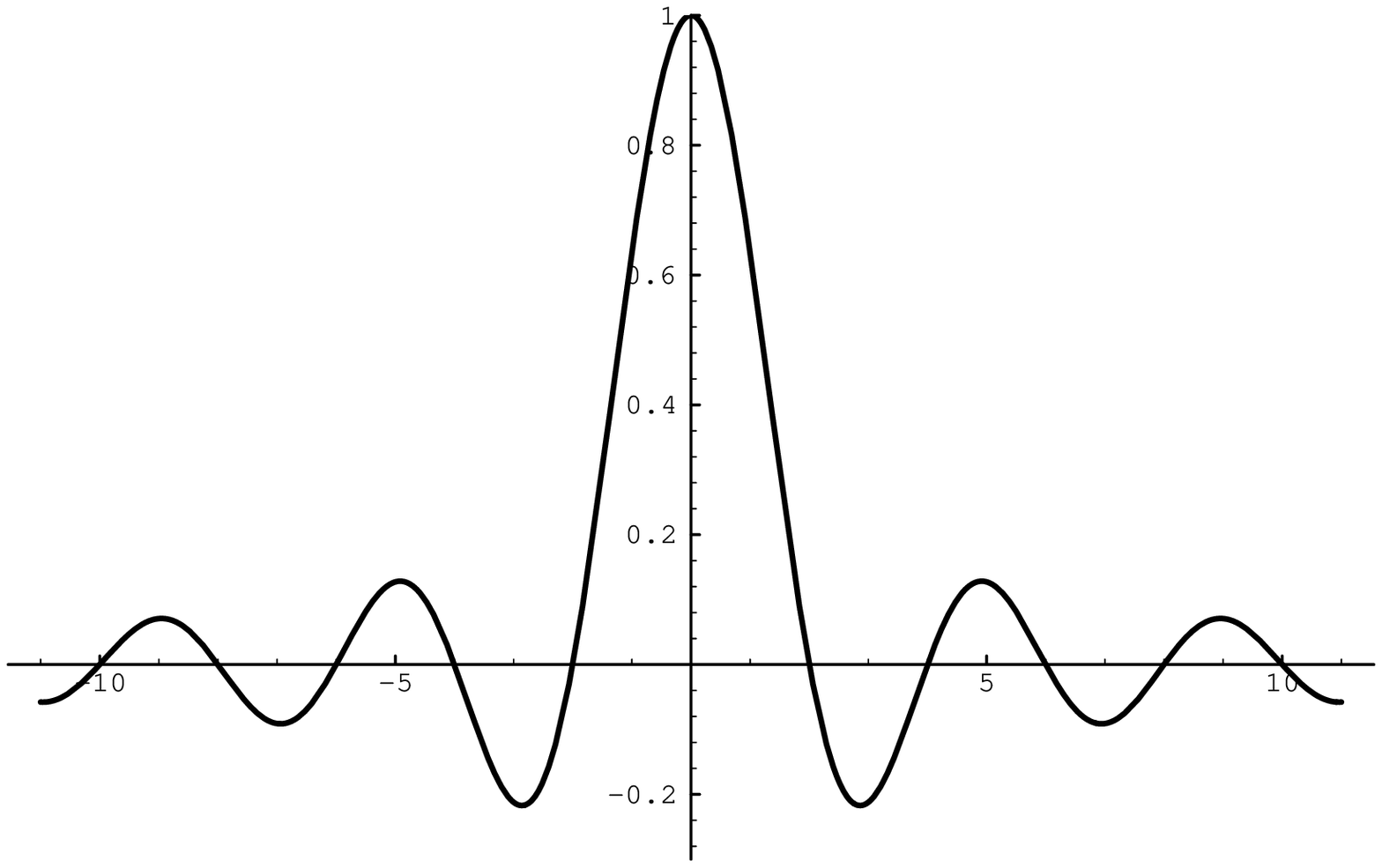}}
\vskip-3.0truecm
\centerline{\small \it Fig2: Plotting 
$\langle \psi_{\lambda}\vert \psi_{\lambda^{\prime}}\rangle $ over
$\lambda-\lambda^\prime$ in units of $\hbar \sqrt{\beta} = 
\Delta x_0$ }
\vskip0.4truecm
Note that this curve is the special case $\Delta p_0 =0$ of the general
curve given as Fig.2 in \cite{ak-jmp-ucr}. Here however, we were also
able to calculate its analytic form. 
\smallskip\newline
The formal position eigenstates are generally no longer
orthogonal. From (\ref{spfp}) we can now however directly 
read off the one-parameter family of diagonalisations
of ${\bf{x}}$. The sets of eigenvectors parametrised by $\lambda \in [-1,1[$
\begin{equation}
\{ \vert \psi_{(2n+\lambda)\hbar\sqrt{\beta}} \rangle
 \vert \quad n \in 
{\:\mbox{\sf Z} \hspace{-0.82em} \mbox{\sf Z}\,} \}
\end{equation}
consist of mutually orthogonal eigenvectors, since
\begin{equation}
\langle \psi_{(2n+\lambda)\hbar\sqrt{\beta}} \vert
\psi_{(2n^\prime+\lambda)\hbar\sqrt{\beta}}\rangle 
= \delta_{n,n^\prime}
\end{equation}
It is not difficult to see that each set is also complete. 
Each of these lattices
of formal ${\bf{x}}$ -eigenvectors has the lattice spacing $2\hbar \sqrt{\beta}$,
which is also $2\Delta x_0$.
\smallskip\newline
Thus in fact there are diagonalisations of the position operator.
One might therefore be tempted to interpret this result such 
that we are now describing physics on lattices in position 
space; for example, compare with the approach 
in ref. \cite{toy1,toy2}.
This is however not the case since the formal
position eigenvectors
 $\vert \psi_{\lambda} \rangle $ are not physical states.
This is because
they are not in the domain of ${\bf{p}}$, which physically
means that they have infinite uncertainty in momentum and 
in particular also infinite energy:
\begin{equation}
\langle \psi_{\lambda}\vert {\bf{p}}^2/2m \vert \psi_{\lambda} \rangle
 =  \mbox{divergent}
\end{equation}
This is in fact also the case for the position eigenvectors of
ordinary quantum mechanics. However, since the uncertainty relation 
holds on every domain of  
symmetric operators ${\bf{x}},{\bf{x}}^2,{\bf{p}},{\bf{p}}^2$ we can in our generalized 
quantum mechanics conclude a much stronger statement:
\medskip\newline
Vectors $\vert \psi \rangle $ that have a well defined uncertainty in 
position $\Delta x_{\vert \psi \rangle }$ which is inside the `forbidden gap'
$$0 \le \Delta x_{\vert \psi \rangle } < \Delta x_0$$ can not have
finite energy.
\medskip\newline
Thus, unlike in ordinary quantum mechanics the formal ${\bf{x}}$-eigenvectors
with their vanishing $x$-uncertainty can now no longer be approximated 
by a series of physical states of 
finite energy where the uncertainty in positions would decrease 
to zero. 
Instead there is now a finite limit to the localizability. 

\section{Recovering information on position}

Generally in quantum mechanics all information on position is encoded 
in the matrix 
elements of the position operator. Matrix elements
 can of course be calculated 
in any basis, e.g. also in the momentum eigenbasis. We now
no longer have any position eigenbasis of 
physical states $\vert x\rangle$
whose matrix elements $\langle x \vert \psi \rangle $
 would have the usual direct physical
interpretation about positions. Nevertheless all information on position
is of course still accessible. To this end let us study the states
which realize the maximally allowed localization.

\subsection{Maximal localization states}
Let us 
explicitly calculate the states $\vert \psi^{ml}_{{\xi}} \rangle $ of 
maximal localization around a position ${\xi}$, i.e. 
states which have the properties
\begin{equation}
\langle \psi^{ml}_{{\xi}}\vert {\bf{x}} \vert \psi^{ml}_{{\xi}}\rangle = {\xi}
\end{equation}
and
\begin{equation}
(\Delta x)_{\vert \psi^{ml}_{{\xi}} \rangle } = \Delta x_0
\end{equation}
We know that $\Delta x_0$ is $\langle {\bf{p}} \rangle$- dependent. 
Recall that the absolutely smallest uncertainty can only be reached 
for $\langle {\bf{p}}\rangle = 0$.

Let us reconsider the (standard) derivation of the uncertainty relation.
For each state in the representation of the Heisenberg algebra
(actually we only need $\vert \psi \rangle $ 
to be in a domain where ${\bf{x}},{\bf{x}}^2,{\bf{p}}$ and ${\bf{p}}^2$
are symmetric) we deduce from the positivity of the norm
\begin{equation}
\| ({\bf{x}} -\langle {\bf{x}} \rangle + \frac{\langle 
[{\bf{x}},{\bf{p}}]\rangle}{2(\Delta p)^2}
(p-\langle p \rangle))\vert \psi \rangle \| \ge 0
\end{equation}
that (note that $\langle [x,p]\rangle$ is imaginary)
\begin{equation}
\langle \psi \vert ({\bf{x}}-\langle {\bf{x}}\rangle)^2 - \left(
\frac{\vert\langle[{\bf{x}},{\bf{p}}]\rangle\vert}{2 (\Delta p)^2}\right)^2
({\bf{p}}-\langle{\bf{p}}\rangle)^2 \vert \psi \rangle  \ge 0
\end{equation}
which immediately implies:
\begin{equation}
\Delta x \Delta p \ge \frac{\vert \langle[{\bf{x}},{\bf{p}}]\rangle\vert}{2}
\end{equation}
It is therefore clear that a state $\vert \psi \rangle $ will obey
$\Delta x \Delta p = \vert \langle[{\bf{x}},{\bf{p}}]\rangle\vert /2$
{\it i.e.} it is on the boundary of the physically allowed region only if
it obeys:
\begin{equation}
({\bf{x}} -\langle {\bf{x}} \rangle + \frac{\langle [{\bf{x}},{\bf{p}}]\rangle}{2(\Delta p)^2}
({\bf{p}}-\langle {\bf{p}} \rangle))\vert \psi \rangle  = 0
\end{equation}         
In momentum space this takes the form of a differential equation
\begin{equation}
\left( i\hbar (1+\beta p^2)\partial_p - \langle {\bf{x}}\rangle
+ i \hbar \frac{1+\beta (\Delta p)^2 + 
\beta \langle {\bf{p}}\rangle^2}{2(\Delta p)^2} 
(p -\langle {\bf{p}}\rangle)\right) \psi(p) = 0
\end{equation}
which can be solved to obtain:
\begin{equation}
\psi(p) = N (1+ \beta p^2)^{-\frac{1+\beta (\Delta p)^2 +\beta \langle
{\bf{p}}\rangle^2}{4\beta (\Delta p)^2}} e^{{\left(\frac{\langle{\bf{x}}\rangle}{
i\hbar\sqrt{\beta}}-\frac{(1+\beta(\Delta p)^2 + \beta\langle {\bf{p}}\rangle^2)
\langle {\bf{p}}\rangle}{2 (\Delta p)^2\sqrt{\beta}}\right) 
\tan^{-1}(\sqrt{\beta} p)}}
\end{equation}
The states of absolutely maximal 
localization can only be obtained for $\langle{\bf{p}}\rangle =0$, see 
eq. (\ref{xmin}). 
We then choose 
the critical momentum uncertainty $\Delta p = 1/\sqrt{\beta}$ to get those
states which are at that point on the curve of 
the uncertainty relation where the minimal
position uncertainty is reached. These states are
\begin{equation}
\psi^{ml}_{\xi}(p) = N (1+ \beta p^2)^{-\frac{1}{2}} {\mbox{ }} e^{-i\frac{
\langle{\bf{x}}\rangle \tan^{-1}(\sqrt{\beta}p)}{\hbar \sqrt{\beta}}}
\end{equation}
where 
\begin{equation}
1 = N N^* \int_{-\infty}^{+\infty} \frac{dp}{(1+\beta p^2)^2} = 
N^2 \frac{\pi}{2\sqrt{\beta}}
\end{equation}
yields their normalization factor $N$.

Thus the momentum space wavefunctions $\psi^{ml}_{{\xi}}(p)$ 
of the states which are maximally 
localized (i.e. $(\Delta x)_{\vert \psi^{ml}_{{\xi}} \rangle }= \Delta x_0$)
around a position ${\xi}$ (i.e. $\langle \psi^{ml}_{{\xi}}\vert {\bf{x}} \vert
\psi^{ml}_{{\xi}}\rangle  = {\xi} $) read:
\begin{equation}
\psi^{ml}_{\xi}(p) = \sqrt{\frac{2\sqrt{\beta}}{\pi}} 
(1+ \beta p^2)^{-\frac{1}{2}} {\mbox{ }} e^{-i\frac{
{\xi} \tan^{-1}(\sqrt{\beta}p)}{\hbar \sqrt{\beta}}}
\end{equation}
These states generalize the plane waves in momentum space or 
Dirac $\delta$- `functions' in position space which would describe maximal
localization in ordinary quantum mechanics. Unlike the latter, the new
maximal localization states are now proper physical states of 
finite energy
\begin{equation}
\langle \psi^{ml}_{{\xi}}\vert \frac{{\bf{p}}^2}{2m} \vert \psi^{ml}_{{\xi}} \rangle =
\frac{2 \sqrt{\beta}}{\pi} \int_{-\infty}^{+\infty} 
\frac{dp}{(1+\beta p^2)^2} \frac{p^2}{2m} = \frac{1}{2m\beta}
\quad .
\end{equation}
Due to the `fuzziness' of space the maximal localization states are 
in general no longer mutually orthogonal:
\begin{eqnarray}
\langle \psi^{ml}_{{\xi}^{\prime}} \vert \psi^{ml}_{{\xi}} \rangle &=&
\frac{2 \sqrt{\beta}}{\pi} \int_{-\infty}^{+\infty}\frac{dp}{(1+\beta p^2)^2}
{\mbox{ }} e^{-i\frac{({\xi} -{\xi}^{\prime})\tan^{-1}(\sqrt{\beta}p)}{\hbar 
\sqrt{\beta}}} \\
 &=& \frac{2 \sqrt{\beta}}{\pi} \int_{-\pi/2}^{+\pi/2} \frac{d\tilde{p}}{
\sqrt{\beta}} \frac{1}{1+\frac{\sin^2(\tilde{p})}{\cos^2(\tilde{p})}}}
e^{-i\frac{{\xi} -{\xi}^{\prime}}{\hbar \sqrt{\beta}} \tilde{p}\\
 &=& \frac{1}{\pi} \left( \frac{{\xi}-{\xi}^{\prime}}{2\hbar\sqrt{\beta}}  
 - \left(\frac{{\xi}-{\xi}^{\prime}}{2\hbar \sqrt{\beta}}\right)^3\right)^{-1}
 \sin\left(\frac{{\xi}-{\xi}^{\prime}}{2\hbar\sqrt{\beta}}\pi\right)
\end{eqnarray}
The poles of the first factor are cancelled by zeros of the sine function.
\vskip-3.0truecm
\epsfxsize=4in \centerline{\epsfbox{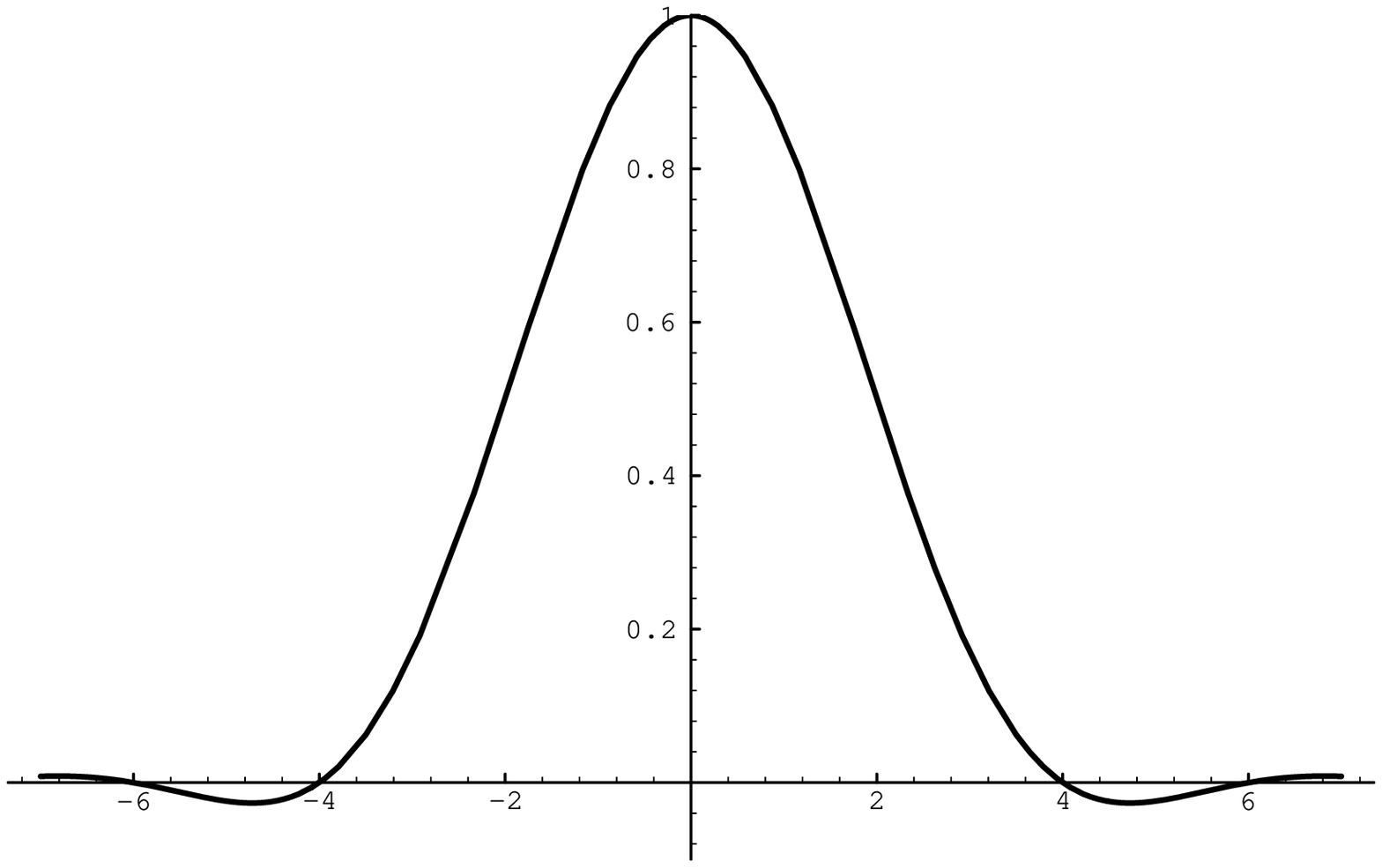}}
\vskip-3.0truecm
\centerline{\small \it Fig3: Plotting 
$\langle \psi^{ml}_{{\xi}^{\prime}} \vert \psi^{ml}_{{\xi}} \rangle$ over
${\xi}-{\xi}^\prime$ in units of $\hbar \sqrt{\beta} = 
\Delta x_0$ }
\vskip0.4truecm
For the width of the main peak note that this curve yields the overlap
of two maximal localization states, each having a standard deviation  
$\Delta x_0$.

\subsection{Transformation to quasi-position wave functions}

While in ordinary quantum mechanics it is often useful to expand
the states $\vert \psi \rangle $ in the position `eigenbasis' $\{ \vert x \rangle\}$ as
$\langle x \vert \psi \rangle$, there are now no 
physical states which would form a position eigenbasis.
Although there is a one parameter family of
${\bf{x}}$-eigenbases, due to the minimal uncertainty gap 
neither of these bases consists of physical
states; furthermore they could not even 
be approximated by physical states of increasing localization.
\smallskip\newline
However we can still project arbitrary states $\vert \phi \rangle$ 
on maximally localized states $\vert \psi^{ml}_{{\xi}} \rangle  $ 
to obtain the probability amplitude for the particle
being maximally localized around the position ${\xi}$.
\smallskip\newline
Let us call the collection of these 
projections $\langle \psi^{ml}_{{\xi}} \vert \phi \rangle $ the state's
`quasi-position wavefunction' $\phi({\xi})$:
\begin{equation}
\phi({\xi}) := \langle \psi^{ml}_{{\xi}} \vert \phi \rangle
\end{equation}
In the limit $\beta \rightarrow 0$ the ordinary position
wave function $\phi({\xi}) = \langle {\xi} \vert \phi \rangle$ is recovered.

The transformation of a state's wavefunction in the momentum 
representation into its quasiposition wave function is of course:
\begin{equation}
\psi({\xi}) = \sqrt{\frac{2\sqrt{\beta}}{\pi}} \int_{-\infty}^{+\infty}
\frac{dp}{(1+\beta p^2)^{3/2}} e^{\frac{i{\xi} \tan^{-1}(\sqrt{\beta} p)}{
\hbar \sqrt{\beta}}} \psi(p)
\label{pxi}
\end{equation}
The quasi-position wavefunction of a momentum eigenstate 
$\psi_{\tilde{p}}(p) = \delta(p-\tilde{p})$, of energy $E=\tilde{p}^2/2m$,
is still a plane wave. However, for its wavelength we obtain a modified
dispersion relation:
\begin{equation}
\lambda(E) = \frac{2\pi \hbar \sqrt{\beta}}{\tan^{-1}(\sqrt{2m\beta E})}
\label{deB}
\end{equation}
The existence of a limit to the precision to which positions can be resolved
manifests itself in the fact that, since the $\tan^{-1}$ is bounded, there
is a nonzero minimal wavelength. The Fourier decomposition of
the quasi-position wave function of
physical states does not contain wavelength components smaller
than 
\begin{equation}
\lambda_0 = 4 \hbar \sqrt{\beta} \quad . \label{deB2}
\end{equation}
Note that, in contrast to ordinary quantum mechanics, since
\begin{equation}
E(\lambda) = \left(\tan\frac{2\pi \hbar 
\sqrt{\beta}}{\lambda}\right)^2\frac{1}{2m\beta}
\end{equation}
quasi-position wavefunctions can no longer have
 arbitrarily fine `ripples', since 
the energy of the short wavelength modes  
diverges as the wavelength approaches the finite value $\lambda_0$.
\smallskip\newline
The transformation (\ref{pxi}) that maps momentum space wave
functions into quasi-position space wave functions is the generalization
of the Fourier transformation and is still invertible. Explicitly, the 
transformation of a quasi-position wavefunction into a momentum space
wave function is easily checked to be:
\begin{equation}
\psi(p) = \frac{1}{\sqrt{8\pi\sqrt{\beta}}\hbar} \int_{-\infty}^{+\infty} 
d{\xi} {\mbox{ }} (1+\beta p^2)^{1/2} {\mbox{ }} e^{-i{\xi}\frac{\tan^{-1}(\sqrt{\beta} p)}{
\hbar \sqrt{\beta}}}{\mbox{ }} \psi({\xi})
\label{transfqptomom}
\end{equation}
Compare also with the generalized Fourier transformation of the discretized
quantum mechanics in \cite{toy2,kw}.
\subsection{Quasi-position representation}

The Heisenberg algebra has a representation on the space
of quasi-position wave functions which we now describe.
Using  eqs. (\ref{sp}) and (\ref{pxi}) the scalar product of 
states in terms of the
quasi-position wavefunctions can be written as
\begin{eqnarray}
\langle \psi \vert \phi \rangle
 & = & \int_{-\infty}^{+\infty} \frac{dp}{1+\beta p^2}
\psi^*(p) \phi(p) \\
  & = & (8\pi\hbar^2\sqrt{\beta})^{(-1)} 
\int_{-\infty}^{+\infty} \int_{-\infty}^{+\infty}
\int_{-\infty}^{+\infty} dp {\mbox{ }} d{\xi} {\mbox{ }} d{\xi}^\prime {\mbox{ }}
e^{i({\xi}-{\xi}^\prime)\frac{\tan^{-1}(\sqrt{\beta}p)}{\hbar 
\sqrt{\beta}}}{\mbox{ }} 
\psi^*({\xi})\phi({\xi}^\prime) \nonumber \label{spq}
\end{eqnarray}
We see from (\ref{transfqptomom}) that the momentum operator is
represented as 
\begin{equation}
{\bf{p}}.\psi({\xi}) = \frac{\tan(-i\hbar\sqrt{\beta}\partial_{{\xi}})}{\sqrt{\beta}} 
\psi({\xi})
\end{equation}        
on the quasi-position wave functions. From the action of 
${\bf{x}}$ on momentum space wave functions 
and eq. (\ref{transfqptomom}) we derive its action on
the quasi-position wave functions:
\begin{equation}
{\bf{x}}.\psi({\xi}) = \left({\xi} + 
\beta \frac{\tan(-i\hbar\sqrt{\beta}\partial_{{\xi}})}{
\sqrt{\beta}}\right) \psi({\xi})
\quad .
\end{equation}

We pause to comment on some important features of the 
quasi-position representation. We found that the position 
and momentum operators ${\bf{x}},{\bf{p}}$ 
can be expressed in terms of the multiplication and differentiation
operators ${\xi}, -i\hbar\partial_{{\xi}}$ which obey the commutation
relations of ordinary quantum mechanics. However, this does not mean 
that we are still dealing with the same space of physical
states with the same properties as in ordinary quantum mechanics.
The scalar product (\ref{spq})
of quasi-position wave functions reduces to the ordinary scalar 
product on position space only for $\beta \rightarrow 0$. Recall also that
the quasi-position wave functions of 
physical states Fourier decompose into wavelengths strictly larger
than a finite minimal wavelength. It is only on 
such physical wave functions that
the momentum operator is defined. 
On general functions of ${\xi}$ the 
power series in the $-i\hbar\partial_{{\xi}}$
which forms the $\tan$ would not be convergent. In addition 
the position operator is not 
diagonalizable in any domain of the symmetric operators ${\bf{x}}^2$ and ${\bf{p}}^2$; 
in particular
the quasi-position representation does 
not diagonalise it. 
The main advantage of the quasi-position representation
is that it has a direct physical interpretation. Recall that $\psi({\xi})$
is the probability amplitude for finding the particle maximally
localized around the position ${\xi}$, i.e. with standard 
deviation $\Delta x_0$.

Let us close with some general remarks on the existence of transformations to
ordinary quantum mechanics and on the significance of the fact that those
transformations are noncanonical.

There are (in $n\ge 1$ dimensions) algebra homomorphisms from generalised Heisenberg
algebras ${\cal{H}}$ generated by operators ${\bf{x}}$ and ${\bf{p}}$ to 
the ordinary Heisenberg
algebra ${\cal{H}}_{0}$ generated by operators ${\bf{x}}_{0}$ and ${\bf{p}}_{0}$.
In one dimension we have {\it e.g.} 
the algebra homomorphism $h: {\cal{H}}\rightarrow {\cal{H}}_{0}$
which acts on the generators as:
$h:  {\bf{p}} \rightarrow {\bf{p}}_{0}$, \quad 
$h: {\bf{x}} \rightarrow {\bf{x}}_{0} + \beta {\bf{p}}_{0} {\bf{x}}_{0} {\bf{p}}_{0}$.
Such mappings $h$ are of representation theoretic interest since 
they induce to any representation $\rho$ of ${\cal{H}}_{0}$ a 
representation $\rho_{h} := \rho \circ h$ of the new
Heisenberg algebra ${\cal{H}}$.

Crucially however, all $h$ are noncanonical. In fact, since 
unitary transformations generally preserve the commutation relations
no representation of ${\cal{H}}$ is unitarily equivalent to any 
representation of ${\cal{H}}_0$. Therefore the set of predictions, 
such as expectation values or transition amplitudes, of a system
based on the new position and momentum operators cannot be matched by 
the set of predictions of any system that is based on 
position and momentum operators obeying the ordinary commutation relations.

\section{A relevant example: The harmonic oscillator}

We would like now to apply the formalism developed thus far
to the interesting case of a linear harmonic oscillator, 
deducing both the energy spectrum and the expression of
 the corresponding eigenfunctions. 
The comparison with the limiting case $\beta =0$ will 
be particularly interesting.\\
{}From the expression for the Hamiltonian:
\begin{equation}
H= \frac{{\bf{p}}^2}{2m} + m \omega^2 \frac{{\bf{x}}^2}{2} 
\label{ham}
\end{equation}
and the representation for ${\bf{x}}$ and ${\bf{p}}$ in the p-space 
reported in Section 3.2 we get the following form for 
the stationary state Schr\"odinger equation:
\begin{equation}
\frac{d^2 \psi(p)}{d p^2} + \frac{2 \beta p}{1 + \beta p^2}
 \frac{d \psi(p)}{d p} +
\frac{1}{(1 + \beta p^2)^2}\left[ \epsilon -
 \eta^2 p^2\right] \psi(p) = 0
\label{schr}
\end{equation}
where we have defined:
\begin{equation}
\epsilon=\frac{2E}{m \hbar^2 \omega^2} 
~;~~~~~~~\eta^2 = \frac{1}{(m \hbar \omega)^2}
\label{epseta}
\end{equation}
and $E$ is the energy.\\
The usual Schr\"odinger equation ($\beta=0$) for
 the linear harmonic oscillator only has one
 singularity at infinity, which is not, however, of the 
Fuchsian kind \cite{smirnov}. In that case
the well-known procedure is to write the solution 
as the product of a decreasing gaussian factor and of a 
new function satisfying an equation, leading to Hermite 
polynomials, where the quadratic term 
$\eta^2 p^2$ is cancelled in the differential equation 
if the gaussian factor is properly chosen. 
{}From eq. (\ref{schr}) we see that the 
introduction of a finite value for $\beta$ 
completely changes the singularity structure in
 the complex plane. Three singular
 points are now present: the usual point
 at infinity as well as  $ p = \pm i/\sqrt{\beta}$. These
 are all regular since the coefficient of the first
 derivative term only behaves as a simple pole in the neighbourhood of
 each singularity; the one in front of the function itself 
contains only double poles\footnote{ We recall
 that in order to study the singular point at infinity
 one should rewrite the equation in terms of the new 
variable $p'=1/p$, shifting the singularity to the origin.}. 
Qualitatively the presence of a minimal 
length  {\it softens} the behaviour of the wave equation at very
 large momenta, transforming the point at infinity into a 
Fuchsian singularity. \\ Equation 
(\ref{schr}) is a Riemann equation whose solution is given in terms
of  hypergeometric functions, which can always be expressed
 in terms of the Gauss hypergeometric series, up to some simple factors.
In order to find the explicit solution it is useful to introduce,
 as usual, a new variable $\zeta$ in terms of which the poles are
 shifted to the reference values $0$, $1$ and $\infty$:
\begin{equation}
\zeta= \frac{1}{2} + i \frac{\sqrt{\beta}}{2} p
\end{equation}
Equation (\ref{schr}) then reads:
\begin{equation}
\frac{d^2 \psi(\zeta)}{d \zeta^2} + \frac{2 \zeta -1 }{\zeta(\zeta-1)} 
\frac{d \psi(\zeta)}{d \zeta} -
\frac{q + r(1 - 2 \zeta)^2}{\zeta^2(\zeta-1)^2} \psi(\zeta) = 0
\label{schrz}
\end{equation}
with: 
\begin{equation}
q= \epsilon/4 \beta~~; ~~~~~~~r= \eta^2/4 \beta^2 
\label{qr}
\end{equation}
whose solution is represented by the Riemann symbol:
\begin{equation}
P \left( \matrix {0&1&\infty&~\cr
\alpha_1&\beta_1&\gamma_1&;\zeta \cr
\alpha_2&\beta_2&\gamma_2&~} \right)
\label{ysol}
\end{equation}
{}From a straightforward computation one gets, for the roots $\alpha_i$, 
$\beta_i$ and $\gamma_i$ \cite{smirnov}:
\begin{eqnarray}
\alpha_1 & = &  -\sqrt{ q+ r}; ~~~ \alpha_2 = \sqrt{q +r} \\ \nonumber
\beta_1 & = &  -\sqrt{ q+ r}; ~~~ \beta_2 = \sqrt{q +r} \\ \nonumber
\gamma_1 & = & \frac{1}{2} ( 1 - \sqrt{1 + 16 r});~~~
\gamma_2 = \frac{1}{2} ( 1 + \sqrt{1 + 16 r})
\label{roots}
\end{eqnarray}
The solution (\ref{ysol}) is simply related to the solution of the 
hypergeometric Gauss equation $F(a,b;c;z)$, for which one of the 
roots for both the singularities at finite distance is zero
\begin{eqnarray}
P \left( \matrix {0&1&\infty&~\cr
\alpha_1&\beta_1&\gamma_1&;\zeta \cr
\alpha_2&\beta_2&\gamma_2&~} \right) & = & 
\zeta^{\alpha_1}(1-\zeta)^{\beta_1} ~
P \left( \matrix {0&1&\infty&~\cr
0&0&a&;\zeta \cr
1-c&c-a-b&b&~} \right) \\ \nonumber
& & \\ \nonumber
 & = &  \zeta^{\alpha_1}(1-\zeta)^{\beta_1} F(a,b;c;\zeta) 
\label{gauss}
\end{eqnarray}
where
\begin{eqnarray}
a & = & \gamma_1 - \alpha_1 - \beta_1 =  \frac{1}{2}
 ( 1 - \sqrt{1 + 16 r}) - 2 \sqrt{q + r} \\ \nonumber
b & = & \gamma_2 - \alpha_1 - \beta_1 =  \frac{1}{2} 
( 1 + \sqrt{1 + 16 r}) - 2 \sqrt{q + r} \\ \nonumber
c & = & 1 - 2 \sqrt{q+r}
\label{abc}
\end{eqnarray} 
We therefore finally get, in terms of the real momentum 
variable $p$, the general form for the solution of the 
equation (\ref{schr}):
\begin{equation}
\psi(p) \propto \frac{1}{(1 + \beta p^2)^{\sqrt{q+r}}} \cdot  
F\left(a,b;c;\frac{1}{2} + i \frac{\sqrt{\beta}}{2} p \right) 
\label{general}
\end{equation}
Since we know that for $\beta=0$ the eigenfunctions are simply
 the product of a Gaussian factor with Hermite polynomials, we now 
look for the solutions for $\beta\ne0$ in the 
cases where the hypergeometric series $F(a,b;c,z)$ reduces to 
a polynomial. This is known to occur 
whenever $a$ or $b$ is a  negative integer: 
\begin{eqnarray}
a=-n & &  ~~  \Rightarrow  \sqrt{q+r} = \frac{1}{2}
 \left(n+ \frac{1}{2} \right)
- \frac{1}{4} \sqrt{1 + 16 r} \label{cond1} \\
b=-n & &   ~~  \Rightarrow  \sqrt{q+r} = \frac{1}{2} 
\left(n+ \frac{1}{2} \right) + \frac{1}{4} \sqrt{1 + 16 r} 
\label{cond2}
\end{eqnarray}
In both cases $F(a,b;c;z)$ becomes a polynomial of degree
 $n$. However if we choose $a=-n$ the wavefunction would
 not have the correct behaviour at infinity and, in particular, 
will not belong to the domain of ${\bf{p}}^2$. From equation (\ref{cond1})
 one has in fact, for large $p$
\begin{equation}
\psi(p) \propto \frac{1}{(1 + \beta p^2)^{\sqrt{q+r}}} \cdot  
F\left(-n,b;c;\frac{1}{2} + i \frac{\sqrt{\beta}}{2} p \right) 
\sim p^{(\sqrt{1+ 16r}-1)/2} 
\label{bad}
\end{equation}
which diverges. Hence 
the condition $b=-n$ yields the energy spectrum
 and the corresponding proper eigenfunctions. In this case  
$\sqrt{q+r}>0$ for any $n$ and for large $p$ the wavefunction behaves as
\begin{equation}
\psi(p) \propto \frac{1}{(1 + \beta p^2)^{\sqrt{q+r}}} \cdot  
F\left(a,-n;c;\frac{1}{2} + i \frac{\sqrt{\beta}}{2} p \right) 
\sim p^{-(\sqrt{1+ 16r}+1)/2} 
\label{good}
\end{equation}
and so is normalizable with respect to the measure $dp/(1 + \beta p^2)$.
 It also belongs to the domain of ${\bf{p}}^2$, as it is immediately checked.
 Note that, for any fixed $n$, the
larger the value of $r$ ({\it i.e.} the smaller the value of
 $\beta$), the more rapid the decay to zero of the wavefunction
 at infinity.
 In particular in the limit $\beta \rightarrow 0$,
 using (\ref{cond2}), 
we recover the usual gaussian
 behaviour of the harmonic oscillator wavefunctions
\begin{equation}
 \lim_{\beta\to0} \frac{1}{(1 + \beta p^2)^{\sqrt{q+r}}}= \exp 
\left(- \frac{\eta^2 p^2}{2} \right)
\label{limgauss}
\end{equation}
 Hence
to each quantum number $n$ there corresponds the eigenfunction
\begin{equation}
\psi_n(p) \propto \frac{1}{(1 + \beta p^2)^{\sqrt{q+r}_n}} \cdot  
F\left(a_n,-n;c_n;\frac{1}{2} + i \frac{\sqrt{\beta}}{2} p \right) 
\label{propeigen}
\end{equation}
where
\begin{equation}
\sqrt{q+r}_n= \frac{1}{2} \left( n + \frac{1}{2} \right) + \frac{1}{4} 
\sqrt{ 1 + 16 r}~~;~~ a_n = -n - \sqrt{1 + 16 r}~~;~~c_n = 1 - 2 \sqrt{q+r}_n
\label{param}
\end{equation} 
For the energy spectrum we obtain 
from (\ref{cond2}) and (\ref{qr}):
\begin{equation}
E_n = \hbar \omega \left( n + \frac{1}{2} \right) \left( \frac{1}{4\sqrt{r}} 
+ \sqrt{ 1 + \frac{1}{16 r}} \right) + \hbar \omega \frac{1}{4\sqrt{r}} n^2
\label{spectrum}
\end{equation}
Notice that the usual spectrum is recovered in the limit
 $\beta \rightarrow 0$ (or $r \rightarrow \infty$); for finite
 $\beta$, the energy levels also depend on the square of the 
quantum number $n$, and asymptotically, for large $n$, they 
grow as $n^2$. In Figure 4 we illustrate for 
comparison the values of the ratio $E_n / \hbar \omega$
for the usual harmonic oscillator and for 
$r=100$.
\vskip-3.0truecm
\epsfxsize=4in \centerline{\epsfbox{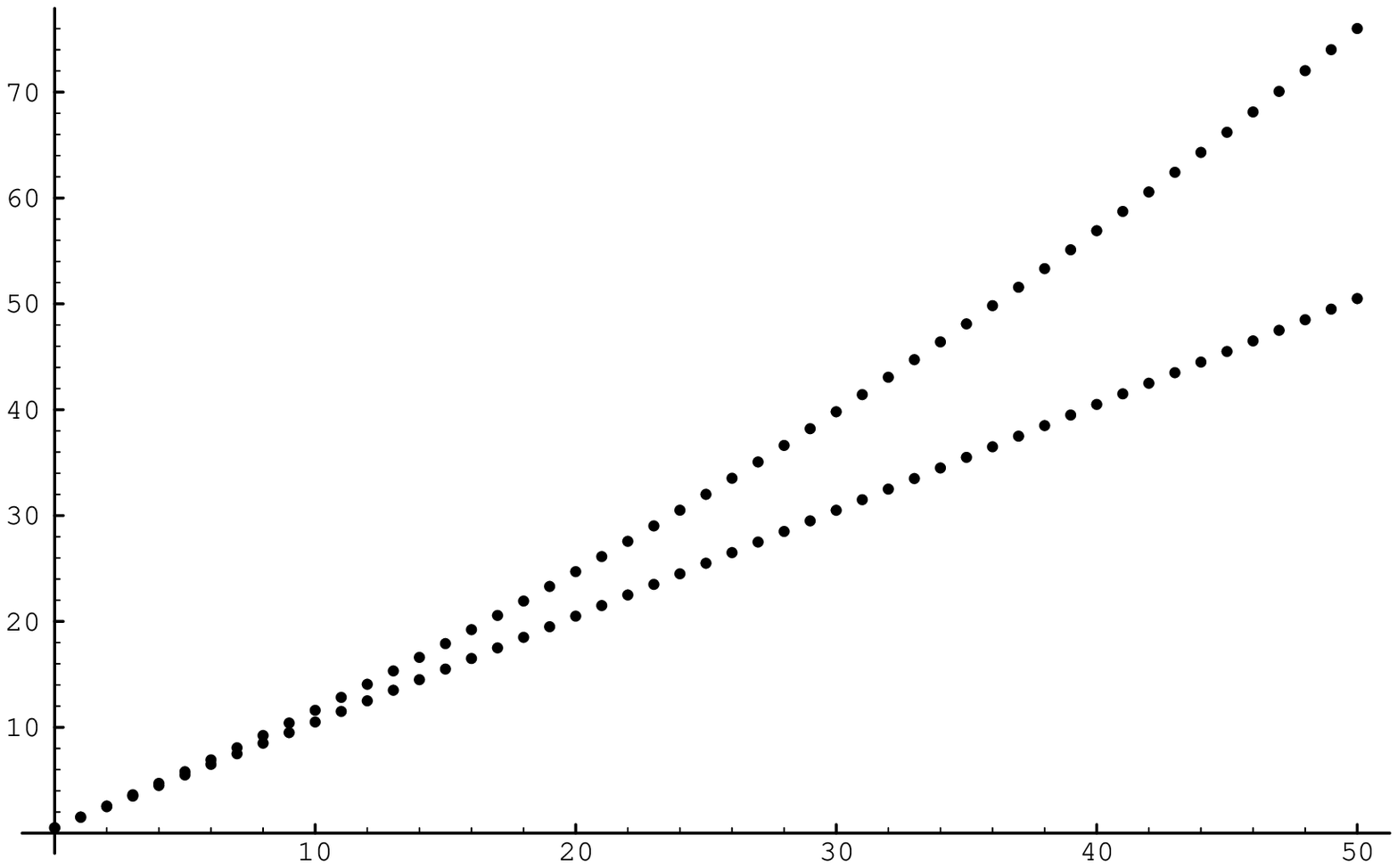}}
\vskip-3.0truecm
\centerline{\small \it Fig4: Comparing 
$E_n/\hbar \omega$ for $r=100$ with the harmonic oscillator}
\centerline{\small \it 
spectrum of ordinary quantum mechanics}
\vskip0.4truecm
We do here not prove the completeness of the set of
 eigenfunctions $\{\psi_n(p)\}$, which is 
quite obvious since the $\psi_n(p)$ adiabatically
 reduce, in the limit $\beta=0$, to the ordinary 
harmonic oscillator eigenfunctions whose 
completeness is known.  


\section{Generalisation to $n$ dimensions}
We now turn to extending the formalism developed in the previous
sections to $n$ spatial dimensions. Our aim is to study an $n$-dimensional
generalization of the framework which still allows
the use of our powerful momentum space representation.
\subsection{Generalised Heisenberg algebra for $n$ dimensions}
A natural generalization of (\ref{cr1dim}) which
preserves the rotational symmetry is:
\begin{equation}
[{\bf{x}}_i,{\bf{p}}_j] = i\hbar \delta_{ij} (1 + \beta {\bf \vec{p}}^2)
\label{nd1}
\end{equation}
It implies nonzero minimal uncertainties 
in each position coordinate.
\smallskip\newline
A more general
situation including nonzero minimal uncertainties in momenta 
has been studied in \cite{ak-jmp-ucr}. 
For the construction of Hilbert space representations for this
general case one cannot work on position space and one has to resort e.g. 
to a generalized Bargmann Fock representation. Here we will specialize
to the situation with nonzero minimal uncertainties in positions only.
\smallskip\newline
We require 
\begin{equation}
[{\bf{p}}_i,{\bf{p}}_j] = 0    \label{nd2}
\end{equation}
which allows us to straightforwardly generalize 
the momentum space representation 
of the previous sections to $n$ dimensions:
\begin{eqnarray}
{\bf{p}}_i.\psi(p) & = & p_i \psi(p) \\
{\bf{x}}_i.\psi(p) & = & i \hbar(1+\beta \vec{p}^2) \partial_{p_i} \psi(p)
\end{eqnarray}
This fixes the commutation relations among 
the position operators. Explicitly we have
\begin{equation}
[{\bf{x}}_i,{\bf{x}}_j]= 2i\hbar\beta({\bf{p}}_i{\bf{x}}_j- {\bf{p}}_j{\bf{x}}_i)
\label{nd3}
\end{equation}
leading  naturally to a `noncommutative geometric' generalization
of position space. Compare also with the approach \cite{doplicher 94/6}.
\smallskip\newline
Note that the generalization to the case  
\begin{equation}
[{\bf{x}}_i,{\bf{p}}_j] = i\hbar \delta_{ij} (1 + f({\bf \vec{p}}^2))
\end{equation}
is straightforward, yielding
\begin{equation}
{\bf{x}}_i.\psi(p) = i\hbar (1+f({\bf \vec{p}}^2))\frac{\partial}{\partial p_i} \psi(p)
\end{equation}
and
\begin{equation}
[{\bf{x}}_i,{\bf{x}}_j] = -2i\hbar f^\prime({\bf \vec{p}}^2) ({\bf{x}}_i {\bf{p}}_j - {\bf{x}}_j {\bf{p}}_i)
\end{equation}
Here we restrict ourselves to the case $f({\bf \vec{p}}^2)= \beta {\bf \vec{p}}^2$.
\smallskip\newline
The operators ${\bf{x}}_i$ and ${\bf{p}}_j$ are 
symmetric on the domain $S_{\infty}$ with respect to the scalar product:
\begin{equation}
\langle \psi \vert \phi \rangle  =  \int_{-\infty}^{+\infty} 
\frac{d^np}{1+ \beta \vec{p}^2} \psi^{*}(p) \phi(p)
\quad .
\end{equation}
The identity operator can be expanded as
\begin{equation}
1  =  
{\int_{-\infty}^{+\infty}\frac{d^np}{1+ \beta \vec{p}^2}}
 \vert p \rangle\langle p \vert 
\end{equation}
and the scalar product of momentum eigenstates is therefore:
\begin{equation}
\langle p \vert p^{\prime} \rangle  = (1+\beta \vec{p}^2) \delta^n(p-p^{\prime})
\end{equation}
While the momentum operators are still essentially self-adjoint,
the position operators are merely symmetric and do not have 
physical eigenstates.
Maximal localization states can again be used to define quasi-position
wave functions. We shall omit a detailed discussion of the 
functional analysis which is
completely analogous to the one-dimensional situation. The 
quasi-position analysis will however be somewhat more involved.
We now focus on the action of the rotation group.

\subsection{Representation of the rotation group}

The commutation relations (\ref{nd1},\ref{nd2},\ref{nd3}) do not break
the rotational symmetry. In fact, 
the generators of rotations can still
be expressed in terms of the position and momentum operators as
\begin{equation}
{\bf{L}}_{ij} := \frac{1}{1+\beta {\bf \vec{p}}^2}({\bf{x}}_i{\bf{p}}_j - {\bf{x}}_j {\bf{p}}_i)
\end{equation}
which in 3 dimensions can be written as
\begin{equation}
{\bf{L}}_k := \frac{1}{1+\beta \vec{{\bf{p}}}^2} \epsilon_{ijk} {\bf{x}}_i{\bf{p}}_j 
\end{equation}
generalizing the usual definition of orbital angular momentum.
Note that $ 1/(1+\beta \vec{{\bf{p}}}^2)$ is an unproblematic bounded operator
acting on the momentum space wave functions as multiplication by
 $ 1/(1+\beta \vec{p}^2)$.
The representation of the generators of rotations
 on momentum wave functions
is
\begin{equation}
{\bf{L}}_{ij}\psi(p) = -i\hbar(p_i\partial_{p_j} - p_j\partial_{p_i})\psi(p)
\end{equation}
where 
\begin{equation}
[{\bf{p}}_i,{\bf{L}}_{jk}] = i\hbar(\delta_{ik}{\bf{p}}_j   -\delta_{ij}{\bf{p}}_k) 
\label{c1}
\end{equation}
\begin{equation}
[{\bf{x}}_i,{\bf{L}}_{jk}] = i\hbar( \delta_{ik}{\bf{x}}_j   -\delta_{ij}{\bf{x}}_k ) 
\label{c2}
\end{equation}
\begin{equation}
[{\bf{L}}_{ij},{\bf{L}}_{kl}] = i\hbar(\delta_{ik}{\bf{L}}_{jl} - \delta_{il}{\bf{L}}_{jk} +
\delta_{jl}{\bf{L}}_{ik} - \delta_{jk}{\bf{L}}_{il})
\label{c3}
\end{equation}
equivalent to ordinary quantum mechanics. However we also have
\begin{equation}
[{\bf{x}}_i,{\bf{x}}_{j}] = -2i\hbar\beta (1+\beta {{\bf \vec{p}}}^2){\bf{L}}_{ij}
\label{c4}
\end{equation}
showing that we have a noncommutative geometry. We therefore have the
uncertainty relations
\begin{equation}
\Delta x_i \Delta p_j \ge \frac{\hbar}{2}\delta_{ij}
 \left(1 + \beta \sum_{k=1}^{n} \left( (\Delta p_k)^2 +
 \langle {\bf{p}}_k\rangle^2
\right) \right)
\label{ucran}
\end{equation}
yielding the minimal uncertainties $\Delta x_0 = \hbar \sqrt{\beta}$ for
each coordinate, and
\begin{equation}
\Delta x_i \Delta x_j \ge \beta \hbar \langle 
(1+\beta{{\bf \vec{p}}}^2){\bf{L}}_{ij}\rangle
\label{ucran2}
\end{equation}
It is straightforward to check that the algebra generated by
(\ref{c1}-\ref{c4}) respects the Jacobi identities.
\smallskip\newline
E.g. in three dimensions one easily verifies that we now have the 
commutation relations:
\begin{equation}
[{\bf{p}}_i,{\bf{L}}_j] = i\hbar \epsilon_{ijk} {\bf{p}}_k 
\end{equation}
\begin{equation}
[{\bf{x}}_i,{\bf{L}}_j] = i\hbar \epsilon_{ijk} {\bf{x}}_k
\end{equation}
\begin{equation}
[{\bf{L}}_i,{\bf{L}}_j] = i\hbar \epsilon_{ijk} {\bf{L}}_k
\end{equation}
\begin{equation}
[{\bf{x}}_i,{\bf{x}}_j] = -2i\hbar\beta (1+\beta {{\bf \vec{p}}}^2)\epsilon_{ijk} {\bf{L}}_k
\end{equation}
where 
\begin{equation}
{\bf{L}}_k.\psi(p) = -i\hbar \epsilon_{ijk} p_i \partial_{p_j} \psi(p)
\end{equation}
is the action of the angular momentum operator on wavefunctions.

\section{Outlook}

The implications of the introduction of a nonzero minimal length in
quantum mechanics are quite profound.  We have shown in the simplest 
non-trivial case that it is no longer possible to spatially localize
a wavefunction to arbitrary precision. The best one can do is consider
the set of maximally localized states as discussed in section IV.
These states have a modified deBroglie relation (\ref{deB}), and
cannot contain wavelength components smaller than the minimal value
$\lambda_0$ given by (\ref{deB2}).  The harmonic oscillator example in
section V shows that the energy levels of a given system can deviate
significantly from the usual quantum mechanical case once energy
scales become comparable to the scale $\sqrt{\beta}$. Although the onset of
this scale is an empirical question, it is presumably set by quantum 
gravitational effects.

We have three possible fields of application in 
mind to extend our work further. First of all,
and most fascinatingly, a minimal length of the form of a minimal 
position uncertainty may indeed describe a genuine feature of 
spacetime, arising directly from gravity. The new kind of 
short distance regularisation therefore has to be implemented 
into quantum field theory. One possibility is to take the general approach
\cite{ak-np2} of employing  the path integral framework, in which
the action functional $S[\phi]$ 
can generally be expressed in terms 
of the scalar product $sp(,)$ and  pointwise multiplication $*$ of fields, where
$x_{\mu}$ and $p^{\nu}$ are formal operators
that act on the fields.
Given a Lagrangian the strategy is to stick to the basis invariant
form of its action functional,
while generalising the commutation relations of the $x_{\mu}$ and $p^{\nu}$.
The modified representation theory then fixes the 
action of the operators, the form of the scalar product and also
largely (though not completely) determines the generalisation of the pointwOAise 
multiplication of fields {\it i.e.} the 
generalisation of local interactions. This then
leads to generalised Feynman rules.
The general studies in \cite{ak-np2} have been carried out in a
generalised Bargmann Fock representation, which is somewhat
difficult to handle. 
The techniques that we have obtained here, in particular the
quasiposition representation,
should now allow a much more detailed examination of the phenomenon
of a minimal position uncertainty in quantum field theory than has 
been possible so far. For example interaction terms, 
which would be slightly nonlocal on 
conventional spacetime, may now be strictly local 
in the sense that (to the extent that short distances can be 
resolved) no 
nonlocality could be observed because of the intrinsic minimal position
uncertainties. A forthcoming paper on this is in progress.

Related to these studies is of course a second possible
field of application, namely the development of a new regularisation
method which does not change the spacetime dimension.

Finally, a quantum theory with minimal uncertainties might be 
useful as an effective theory of nonpointlike particles. These
could, for example, be strings since our formalism includes 
the uncertainty relation obtained in string
theory, see {\it e.g.} \cite{maggiore}. 
In addition, this suggests investigating whether 
our formalism may also be used to effectively describe compound particles 
such as mesons in situations where their nonzero size matters but where
details of their internal 
structure does not contribute. The parameter
$\beta$ would then of course not be at the Planck scale but at a scale 
that would relate the `extension' of the particle to the minimal uncertainty.
Furthermore, since quasi-particles and collective excitations cannot 
be localised to arbitrary precision, there 
might be a possibility of including some of the 
first order dynamical behavior of such systems in the kinematical effects
of generalised uncertainty relations, thereby improving (or at least
simplifying) the effective description.

\end{document}